%%%%%%%%%%%%%%%%%%%%%%%%%%%%%%%%%%%%%%%%%%%%%%%%%%%%%%%%%%%%%%%%%%%%%%%%%%%
%%
%%            q-IDENTITIES AND AFFINIZED PROJECTIVE VARIETIES
%%                       II.  Flag varieties
%%
%%                 Peter Bouwknegt and Nick Halmagyi
%%
%%
%%  ADP-99-2/M77                    
%%  math-ph/9903033                                           March, 1999
%%
%%%%%%%%%%%%%%%%%%%%%%%%%%%%%%%%%%%%%%%%%%%%%%%%%%%%%%%%%%%%%%%%%%%%%%%%%%%
\input amstex
\input epsf

\documentstyle{amsppt}
\magnification=1200
\hfuzz=35pt

\pagewidth{16.5truecm}
\pageheight{22truecm}

% -------------------------------------------------------------------------
% \input amsmac

%
%  Greek abbreviations 
%
\def\al{\alpha}
\def\be{\beta}
\def\ga{\gamma}    
\def\de{\delta}    \def\De{\Delta}
\def\ep{\epsilon}

\def\la{\lambda}   \def\La{\Lambda}

\def\si{\sigma}    
\def\ta{\tau}
  
      \def\Ph{\Phi}    
\def\ch{\chi}

%
%  boldface abbreviations 
%

\def\bI{{\bold I}}

\def\bS{{\bold S}}
\def\bV{{\bold V}}

%
% Calligraphic abbreviations 
%

\def\cP{{\Cal P}}

%

%----------------------------------------------------------------------------
% SOME ADDITIONAL MATH SYMBOLS
%
%\newsymbol\box 1003

\def\CC{{\Bbb C}}
\def\ZZ{{\Bbb Z}}

\def\PP{{\Bbb P}}

\def\bfg{{\frak g}}

\def\hg{{\widehat{\frak g}}}
\def\whg{\hg}

\def\eql{~=~}

\def\qeii{QEI$\!$I}
\def\wh#1{\widehat{#1}} \def\wt#1{\widetilde{#1}}
\def\deg{{\text{\rom{deg}}}}

\def\qn#1{(q)_{#1}}
\def\half{\textstyle{1\over2}}
 
\def\shskip{\medskip}
\def\qbinom#1#2{\left[ \matrix {#1} \\ {#2} \endmatrix \right] }
\def\p{\partial}
\def\sym{\text{Sym}} \def\symb{\sym^{\bullet}}
\def\xd#1{x_{#1}} \def\dxd#1{ {\p\over \p x_{#1}}}
\def\xu#1{x^{#1}} 
\def\bone{{\bar 1}} \def\btwo{{\bar 2}} \def\bthr{{\bar 3}}

\def\bG{\text{\bf G}} \def\bM{\text{\bf M}} \def\ba{\text{\bf a}}
\def\bA{\text{\bf A}} \def\id{\text{\bf 1}}
\def\bu{\text{\bf u}}
\def\bm{\text{\bf m}}

\def\PP{{\Bbb P}}

\def\mywedge{{\textstyle{\bigwedge}}}
\def\pars{\smallskip}
% -------------------------------------------------------------------------
%\leftline{\epsffile{y2k.eps}}
\rightline{ADP-99-2/M77}
\rightline{\tt math-ph/9903033}\bigskip

\topmatter
\title $q$-identities and affinized projective varieties\\
II.\ Flag varieties\endtitle
\rightheadtext{$q$-identities and affinized projective varieties, II}
\author Peter Bouwknegt and Nick Halmagyi\endauthor
%\affil \endaffil
\address Department of Physics and Mathematical Physics, University of 
Adelaide, Adelaide SA~5005, AUSTRALIA\endaddress
%\curraddr \endcurraddr
\email pbouwkne\@physics.adelaide.edu.au \endemail
\address Department of Physics and Mathematical Physics, University of 
Adelaide, Adelaide SA~5005, AUSTRALIA\endaddress
\curraddr Department of physiology, University of Sydney NSW~2006,
AUSTRALIA\endcurraddr
\email nhalmagy\@physiol.usyd.edu.au \endemail
\date  \enddate
\dedicatory Dedicated to the memory of Prof.\ H.S.~Green\enddedicatory
%\thanks Supported by a \qeii\ Fellowship from the 
%Australian Research Council \endthanks
%\translator \endtranslator
%\keywords projective variety, quasi-particle, q-identity\endkeywords
%\subjclass \endsubjclass
\abstract 
In a previous paper we defined the concept of an affinized projective
variety and its associated Hilbert series.  We computed the Hilbert 
series for varieties associated to quadratic monomial ideals.
In this paper we show how to apply these results to affinized flag
varieties.  We discuss various examples and conjecture a correspondence
between the Hilbert series of an affinized flag variety and a modified
Hall-Littlewood polynomial.  We briefly discuss the application of these
results to quasi-particle character formulas for 
affine Lie algebra modules.
\endabstract
\endtopmatter
% -------------------------------------------------------------------------
\document

% -------------------------------------------------------------------------
\head 1. Introduction \endhead

In a previous paper \cite{Bo}, one of us introduced the concept of an
affinized projective variety and the associated Hilbert series of its
coordinate ring.  It was shown how $q$-identities naturally arise from
different ways of computing this Hilbert series.  This was made
explicit in the case of projective varieties defined by quadratic
monomial ideals, where, on the one hand, an explicit basis for the
coordinate ring of the associated affinized projective variety led to
an expression of the Hilbert series, while, on the other hand, an
algorithm was described how to obtain an alternating sum formula
(resembling an `affinization' of Taylor's resolution for monomial
ideals) for the same Hilbert series. \pars

In the present paper we will show how one can apply the results of 
\cite{Bo} to compute the Hilbert series of an affinized flag variety.
The ideal that defines the coordinate ring of
a flag variety, albeit quadratic, is not a monomial ideal.  We will 
show, however, how one can in principle construct a projective 
variety defined by a quadratic monomial ideal which has the {\it same}
Hilbert series.  This involves two ideas.  The first is to consider 
the (monomial) ideal of leading terms $\langle \text{LT}(I)\rangle$
instead of $I$.  The second idea is to remove the non-quadratic monomials
by the addition of extra variables.  We carry out this program in 
several examples. \pars

The Hilbert series of a flag variety based on a group $G$, as
well as the Hilbert series of its affinization, are naturally 
characters of $\text{Lie}\,G = \bfg$.  In fact, upto a 
trivial factor, the Hilbert series
of the affinized flag variety can be interpreted as a $q$-deformation
of the character of a tensor product of $\bfg$-modules.
One of the main results of this paper is a conjectured correspondence 
between the Hilbert series of an affinized flag variety and another such
$q$-deformation, namely the modified Hall-Littlewood polynomial (or Milne
polynomial).
Modified Hall-Littlewood polynomials are related in various ways to 
characters of affine Lie algebras.  Our results thus provide new, explicit,
expressions for these characters.  In fact, this was the main motivation 
for this work.  A relation between the characters of affine Lie algebras
and the geometry of infinite dimensional flag varieties was also
pointed out and explored in \cite{FS1,FS2}.
\pars

The affine Lie algebra characters are naturally obtained in what is known
as the `Universal Chiral Partition Function' (UCPF) form,
which is conjectured to be a universal form for the (chiral) characters
of any two dimensional conformal field theory \cite{BM}, and is 
closely related to a description of that conformal field theory in
terms of quasi-particles.  The results of this paper further support
the validity of this conjecture, at least in the case of conformal 
field theories based on affine Lie algebras (WZW models).
In a separate paper \cite{BCR} we will
explore the results of this paper in connection with the exclusion 
statistics (cf.\ \cite{Ha,BS4,GS} and
referencse therein) satisfied by these quasi-particles.
The example of $\frak{so}_5$, treated in detail in this paper, is 
particularly relevant with regards to possible applications to the 
quasi-particles (`non-abelian electrons') in $SO(5)$ superspin regimes 
of strongly correlated electrons on a two-leg ladder \cite{BS3}.
\pars

This paper is organized as follows.  In section 2 we recall the 
definition of the homogeneous coordinate ring of a 
(finite dimensional) flag variety and show how one can, in principle,
compute the ideal of quadratic relations.  We illustrate this in a few
examples.
In section 3 we introduce affinized flag varieties and outline a procedure
to compute the Hilbert series of such a variety.  
In section 4 we apply the procedure in a few examples, namely,
$\frak{sl}_n$, $n=2,3,4$, and $\frak{so}_{2n+1}$, $n=2,3$, and
in section 5 we comment on the relation between the Hilbert series 
of an affinized flag variety and modified Hall-Littlewood polynomials
as well as characters of integrable highest weight 
modules of affine Lie algebras.
We also comment on the application of our results to theories of 
quasi-particles, their exclusion statistics and the UCPF.\pars

Throughout this paper we will use the notation of \cite{Bo}.

% -------------------------------------------------------------------------
\head 2. Flag varieties \endhead

% -------------------------------------------------------------------------
\subhead 2.1. Generalities \endsubhead\shskip

Most of the material in this section is quite standard.  We refer
to \cite{FH,Fu} for detailed expositions. \pars

Let $\frak g$ be a (complex) finite dimensional simple Lie algebra of
rank $\ell$.  Let $\{\al_i\}_{i=1}^\ell$ and $\{\La_i\}_{i=1}^\ell$
denote the simple roots and fundamental weights of $\frak g$, respectively.
Let $L(\La_i)$, $i=1,\ldots,\ell$, denote the 
finite dimensional irreducible representation of $\frak g$ with highest 
weight $\La_i$ and dimension $D_i = \text{dim}\, L(\La_i)$.  For any
$\frak g$-module $V$, we denote by $\sym^M\,V$ the symmetrized tensor
product of $M$ copies of $V$, and by $\symb\,V = \oplus_{M\geq0}
\sym^M\,V$ the symmetric algebra on $V$.   The module 
$$
\sym^{M_1} L(\La_1) \otimes \ldots
\otimes \sym^{M_\ell} L(\La_\ell) \,, \tag{2.1}
$$
is completely reducible, for any choice of $M_i\geq0$, and contains 
$L(M_1\La_1+\ldots+M_\ell\La_\ell)$ as a submodule.  Upon introducing
coordinates $x^{(i)}_a$, $a=1,\ldots,D_i$, with respect to an 
orthonormal basis $v_a^{(i)}$ for each $L(\La_i)$, $i=1,\ldots,\ell$, 
we can identify 
$$ \align
\bS & ~\cong~ \symb\,L(\La_1) \otimes \ldots\otimes \symb\, L(\La_\ell)\,,\\
\bS_{(M_1,\ldots,M_\ell)} & ~\cong~ \sym^{M_1} L(\La_1) \otimes \ldots
\otimes \sym^{M_\ell} L(\La_\ell) \,, \tag{2.2}\endalign
$$
where $\bS \equiv \CC[x_a^{(i)}]_{a=1,\ldots,D_i}^{i=1,\ldots,\ell}$ 
and $\bS_{(M_1,\ldots,M_\ell)}$
denotes the subspace of polynomials in $\bS$, homogeneous of degree
$M_i$ in the $\{x^{(i)}_a\}_{a=1,\ldots,D_i}$.  The Lie algebra
$\frak g$ is realized on $\bS_{(M_1,\ldots,M_\ell)}$ in terms of linear 
differential operators.  Note, furthermore,
$$
\text{dim}\,\bS_{(M_1,\ldots,M_\ell)} \eql
\prod_{i=1}^\ell \ \binom{M_i + D_i -1}{D_i-1}\,. \tag{2.3}
$$
Of course, the main problem is to give a concrete description for how the 
irreducible module $L(M_1\La_1+\ldots+M_\ell\La_\ell)$ is contained in
$\bS_{(M_1,\ldots,M_\ell)}$.  The solution is remarkably simple. 
By a theorem of Kostant (see, e.g., \cite{LT}) we have 
that $\bS$, as a $\frak g$-module, contains an ideal $I$ generated by
quadratic relations such that 
$$
L(M_1\La_1+\ldots+M_\ell\La_\ell) ~\cong~ \bS(V)_{(M_1,\ldots,M_\ell)}\,,
\tag{2.4}
$$
where $\bS(V)=\bS/I$.
Furthermore, it turns out that $\bS(V)$ is precisely the 
homogeneous coordinate ring of the (complete) flag variety of $\frak g$,
i.e., the variety $V = \bV(I)$, corresponding to $I\subset 
\CC[x_a^{(i)}]$, is isomorphic to $G/B$ where $G$ is the (complex) Lie group 
such that $\text{Lie}\,G=\frak g$ and $B$ is a Borel subgroup of $G$.
More generally, omitting some of the variables $x_a^{(i)}$ for certain
$i$ (or, equivalently, taking $M_i=0$ for some $i$), corresponds to
the homogeneous coordinate ring of a partial flag variety $G/P$ where 
$P$ is a parabolic subgroup of $G$.\pars

The ideal of quadratic relations $I$ can be determined, in principle,
by analyzing the tensor products $L(\La_i)\otimes  L(\La_j)$ (see
the examples in sections 2.2 and 2.3).  It appears that explicit results
are known only for $\frak g=\frak{sl}_n$ (or $\frak{gl}_n$) and some
low rank cases. \pars

Besides the grading by the multi-degree $\bM=(M_1,\ldots,M_\ell)$, the 
$\bS$-module $\bS(V)$ is also graded by the $\frak g$-weight.  Denote
by $\bS(V)_{(\bM;\la)}$ the space of homogeneous polynomials in 
$\bS(V)$ of multi-degree $\bM$ 
and $\frak g$-weight $\la$.  The Hilbert function 
of $\bS(V)$ is then defined as the (formal) $\frak g$-character
$$
h_V(\bM) \eql \sum_{\la} \ \text{dim}\,
\bS(V)_{(\bM;\la)} \, e^\la\,, \tag{2.5}
$$
or, by (2.4), as the character of the irreducible module $L(M_1\La_1+\ldots
+M_\ell\La_\ell)$.\pars

As discussed in \cite{Bo}, the Hilbert function $h_V(\bM)$ can be computed
either by constructing an explicit basis for the module $\bS(V)$ or 
by applying the Euler-Poincar\'e principle to a free resolution of $\bS(V)$
$$
0 @>>> F^{(\nu)} @>d_\nu>> \ldots @>d_3>> F^{(2)} 
@>d_2>> F^{(1)} @>d_1>> F^{(0)} \cong \bS @>>> \bS(V) @>>> 
0\,, \tag{2.6}
$$
where the maps $d_i$ intertwine with the 
action of $\frak g$.  Here, $F^{(j)} = \oplus_k 
\bS(-\ba^{(j)}_{k};-\mu^{(j)}_{k})$ 
for some set of vectors $\ba^{(j)}_{k} = 
((a^{(j)}_{k})_1,\ldots,(a^{(j)}_{k})_\ell)$ with positive integer 
coefficients, and integral weights $\mu^{(j)}_{k}$.  In fact,
because of the $\frak g$ structure, the weights $\mu^{(j)}_{k}$ nicely
combine into weights of $\frak g$-modules, i.e., we can write
$F^{(j)} = \oplus_k (\bS(-\ba^{(j)}_{k}) \otimes V_k^{(j)})$ for some set of 
(finite dimensional) $\frak g$-modules $V_k^{(j)})$.
Furthermore, 
as in \cite{Bo}, for any module $M$ 
we have denoted by $M(\ba;\mu)$ the same module with multi-degree and weight
shifted by $\ba$ and $\mu$, respectively.  Application of the 
Euler-Poincar\'e principle to (2.6) yields the following alternating sum
formula for the character $h_V(M_1,\ldots,M_\ell)$ of the irreducible module
$L(M_1\La_1+\ldots+M_\ell\La_\ell)$ 
$$
h_V(M_1,\ldots,M_\ell) \eql \sum_{j}\ (-1)^j 
\sum_k\ h_{\bS}(M_1-(a^{(j)}_{k})_1,
\ldots,M_\ell -(a^{(j)}_{k})_\ell)\, \ch_{V_k^{(j)}}\,,\tag{2.7}
$$
where $\chi_V$ denotes the (formal) character of the $\frak g$-module $V$ and 
$$
h_{\bS}(M_1,\ldots,M_\ell) \eql  \sum_\la \text{dim}\,
\bS_{(M_1,\ldots,M_\ell;\la)} e^\la 
\eql  \sum_{ \sum_a m_a^{(i)} = M_i } 
e^{ \sum_{i,a} m_a^{(i)} \la_a^{(i)} } \,,\tag{2.8}
$$
denotes the character valued Hilbert function of $\bS=\CC[x_a^{(i)}]$.  
The weights 
of $L(\La_i)$ are denoted by $\la_a^{(i)}$, $a=1,\ldots,D_i$.\pars

The existence of a (finite) free
resolution (2.6) is guaranteed by Hilbert's syzygy theorem, but 
explicit descriptions are known apparently only in special examples.
Some examples will be discussed in sections 2.2 and 2.3.  Based on these
and other examples we have 
\proclaim{Conjecture 2.1} Consider the minimal resolution (2.6) of $\bS(V)$. 
Let $\nu$ be the length of this resolution, $D = \text{dim}\, G/B$,
$\ell = \text{rank}\,{\frak g}$ and $D_i=\text{dim}\,L(\La_i)$,
then
\item{i.}   $\sum_{i=1}^\ell D_i = D + \ell + \nu$
\item{ii.} The minimal
   resolution is symmetric under simultaneous interchange of 
  $\bS(a_1,\ldots,a_\ell;\mu) \leftrightarrow
   \bS(2-D_1-a_1,\ldots,2-D_\ell-a_\ell;-\mu)$ 
  (or, equivalently, $\bS(a_1,\ldots,a_\ell)\otimes V \leftrightarrow
  \bS(2-D_1-a_1,\ldots,2-D_\ell-a_\ell)\otimes V^*$, where $V^*$ is the 
  $\frak g$-module contragredient to $V$), and reversal of all arrows.
\endproclaim\medskip

In particular, 
since $F^{(0)} = \bS = \bS(0,\ldots,0;0)$ it follows from ii.\ that 
$F^{(\nu)} \cong \bS(2-D_1,\ldots,2-D_\ell;0)$.\pars

In the remainder of this section we will explicitly go through some examples
in preparation for the affinization of this construction which will
be discussed in sections 3 and 4.

% -------------------------------------------------------------------------
\subhead 2.2. Example: $\frak{sl}_n$ \endsubhead\shskip

Consider the Lie algebra $A_{n-1} \cong
\frak{sl}_n$ of rank $\ell=n-1$.  In terms of 
an overcomplete basis $\{ \ep_i \}_{i=1}^n$
of $\CC^{n-1}$, satisfying 
$$
(\ep_i, \ep_j) \eql \de_{ij} - \textstyle{{1\over n}} \,, \qquad
\sum_{i=1}^n \ep_i \eql 0\,,\tag{2.9}
$$
the simple roots and fundamental weights of $\frak{sl}_n$ can be written as
$$ 
\al_i  \eql \ep_i - \ep_{i+1} \,,\qquad 
\La_i  \eql \ep_1 + \ldots + \ep_i \,, \qquad i=1,\ldots,n-1\,.
\tag{2.10}
$$
The irreducible finite dimensional representation $L(\La_i)$,
of dimension $D_i = \binom{n}{i}$, has weights 
$$
\ep_{i_1} + \ldots + \ep_{i_k}\,, \qquad 1\leq i_1 < \ldots < i_k \leq n\,.
\tag{2.11}
$$
Let us denote the 
coordinate corresponding to the vector of weight 
$\ep_{i_1} + \ldots + \ep_{i_k}$ by $x_{i_1\ldots i_k}$.  For convenience
we extend the definition of $x_{i_1\ldots i_k}$ to
arbitrary sequences $1\leq i_1,\ldots,i_k\leq n$ by anti-symmetry, and raise 
and lower indices by means of the $\ep$-tensor, i.e.,
$$
x^{i_1\ldots i_{n-k}} \eql {\textstyle{1\over k!}}\
\ep^{i_1\ldots i_{n-k}j_1\ldots j_k}
  x_{j_1\ldots j_k} \,. \tag{2.12}
$$
We have a realization of $\frak{gl}_n$ on $\bS=
\CC[x_{i_1},\ldots, x_{i_1\ldots i_{n-1}}]$ in terms of linear
differential operators, given by
$$
e_{ij} \eql \xd{i} \dxd{j} + \sum_k \xd{ik}\dxd{jk} + \ldots
  + \sum_{k_1<\ldots<k_{n-2}} \xd{ik_1\ldots k_{n-2}} 
  \dxd{jk_1\ldots k_{n-2}} \,. \tag{2.13}
$$
That is, the operators (2.13) satisfy the defining relations of $\frak{gl}_n$
$$
[e_{ij},e_{kl}] \eql \de_{jk} e_{il} - \de_{il} e_{kj} \,.\tag{2.14}
$$
Henceforth, we consider $\bS$ as an $\frak{sl}_n$-module.  As argued before,
the $\frak{sl}_n$-module $\bS_{(M_1,\ldots,M_{n-1})}$ contains $L(\La) \equiv
L(M_1\La_1+\ldots+M_{n-1}\La_{n-1})$ as a submodule.  In particular, 
the highest weight vector of $L(\La)$ is given by 
$x_1^{M_1} x_{12}^{M_{2}}\ldots x_{12\ldots n-1}^{M_{n-1}}$. \pars

It is well-known (see, e.g.,
\cite{FH,Fu}) that the ideal $I$ of quadratic relations is generated 
by the polynomials 
$\{ \si_{i_2\ldots i_k}^{j_1\ldots j_{n-l-1}} \ | \ 1\leq k\leq l<n\}$,
where
$$
\si_{i_2\ldots i_k}^{j_1\ldots j_{n-l-1}} \eql
  \ep^{i_1k_1\ldots k_l j_1\ldots j_{n-l-1}} x_{i_1i_2\ldots i_k}
  x_{k_1k_2\ldots k_l}\,, \tag{2.15}
$$
Note that the invariant subspace of polynomials 
$\si_{i_2\ldots i_k}^{j_1\ldots j_{n-l-1}}$, 
for fixed $1\leq k \leq l < n$, is isomorphic to 
$L(\La_{k-1}) \otimes L(\La_{l+1})$ and will thus, in general, be
reducible under $\frak{sl}_n$. \pars

\remark{Remark 2.1}  We briefly comment on the identification of the 
coordinate ring defined above with that of a flag variety in the 
case of $\frak{sl}_n$ (see \cite{Fu} for more details).
Let $E= \CC^n$.  Fix a sequence $n\geq d_s > \ldots >d_2 >d_1\geq0$.
The (partial) flag variety $\text{Fl}^{d_1\ldots d_s}(E)$ is the set
of flags 
$$
\{ E_s \subset \ldots \subset E_2 \subset E_1 \subset E\ |\ 
\text{codim}\,E_i=d_i,\, 1\leq i \leq s\}\,. \tag{2.16}
$$
The group $G=GL(E)=GL(n,\CC)$ acts transitively on 
$\text{Fl}^{d_1\ldots d_s}(E)$.  
Now, let $\{e_i\}_{i=1}^n$ be a basis of $E$ and let 
$F^{(0)} =\{
E_s^{(0)} \subset \ldots \subset E_2^{(0)} \subset E_1^{(0)} \subset E\}$
be the flag defined by $E_i^{(0)} = \langle e_{d_i+1}, e_{d_i+2},\ldots,
e_n\rangle$.  Furthermore, let $P$ be the parabolic subgroup of $GL(E)$ 
fixing the flag $F^{(0)}$, i.e.,
$$
P \eql \{ g\in GL(E)\ | \ g(E_i) \subset E_i, 1\leq  i \leq s\}\,, \tag{2.17}
$$
then we can identify $\text{Fl}^{d_1\ldots d_s}(E) = G/P$.
Specifically, note that a matrix in $P$ has invertible matrices in blocks of
sizes $d_1,d_2-d_1,\ldots,d_s-d_{s-1},n-d_s$ down the diagonal, with 
arbitrary entries below these blocks
and vanishing entries above.  For the complete flag variety 
$\text{Fl}^{1,2,\ldots,n}(E)$ we have $P=B$ where $B$ is a Borel subgroup
of $G$.  We have a natural embedding, the so-called Pl\"ucker
embedding,
$$
\text{Fl}^{d_1\ldots d_s}(E) ~\hookrightarrow~ \PP(\mywedge{}^{d_1}E) \times
\ldots\times \PP(\mywedge{}^{d_s}E)\,. \tag{2.18}
$$
Since $\bigwedge^p E$ has a basis $e_{i_1}\wedge\ldots\wedge e_{i_p}$, we have
a natural set of homogeneous coordinates 
$x_{i_1\ldots i_p}$ for $\PP(\bigwedge^{p}E)$.  The relations (2.15)
are precisely the ``Pl\"ucker relations'' for the embedding (2.18).
\endremark\medskip

For $\frak{sl}_n$ the (minimal) resolution (2.6) of
$\bS(V) = \bS/I$ does not seem to be known for general $n$.  
We discuss some examples for small $n$.

% -------------------------------------------------------------------------
\subsubhead 2.2.1. $\frak{sl}_2$ \endsubsubhead \shskip

For $\frak{sl}_2$ we have a single fundamental 
representation $L(\La_1)$ of dimension $D_1 = 2$.  Moreover,
$\sym^2 L(\La_1) \cong L(2\La_1)$, so 
$\bS=\CC[x_1,x_2]$ while the ideal $I$ is trivial.
We immediately conclude that the character of the irreducible (spin-$M/2$)
module $L(M\La_1)$ is given by (cf.\ (2.7))
$$
h(M) \eql \sum_{m_1+m_2=M} \ e^{m_1\ep_1+m_2\ep_2}
\eql \sum_{m_1+m_2=M} \ e^{(m_1-m_2)\ep_1}  \,,\tag{2.19}
$$
as one may readily verify.

% -------------------------------------------------------------------------
\subsubhead 2.2.2. $\frak{sl}_3$ \endsubsubhead \shskip

We have two fundamental representations $L(\La_i)$, $i=1,2$,
of dimension $D_1=D_2=3$, hence $\bS=\CC[ x_i ,x^i]_{i=1,2,3}$.  From 
$$ \align
\sym^2 L(\La_1)   &  ~\cong~ L(2\La_1)\,, \\
L(\La_1) \otimes L(\La_2)  &  ~\cong~ L(\La_1+ \La_2) \oplus L(0) \,,\\
\sym^2 L(\La_2) &  ~\cong~ L(2\La_2) \,,\tag{2.20} \endalign
$$
it follows that $I$ is generated by a single ($\frak{sl}_3$-singlet)
generator $\si$.  Explicitly,
$$
\si \eql \xd{i} \xu{i} \eql \half \ep^{ijk} \xd{i} \xd{jk} \,.\tag{2.21}
$$
The resolution of $\bS(V) = \bS/I$ is obviously given by
$$
0 @>>> F^{(1)} @>d_1>> F^{(0)} @>>> \bS/I @>>> 0\,, \tag{2.22}
$$
where $F^{(1)} \cong \bS(-1,-1)$ and $d_1:F^{(1)}\to F^{(0)}$ is 
defined by
$$
d_1\,:\, e^{(1)} ~\mapsto~ \si\, e^{(0)} \,, \tag{2.23}
$$
where $e^{(i)}$ is the generator of $F^{(i)}$.  See also Fig.\ 2.1.

\bigskip
\centerline{\epsffile{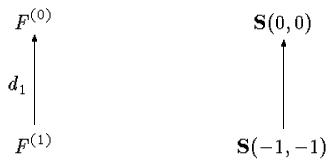}}
\bigskip
\centerline{Fig.\ 2.1.\ Resolution of $\bS(V)$ for $\frak{sl}_3$}
\bigskip

The resolution (2.22) leads to the following well-known expression for the
character $h_V(M_1,M_2)$ of the irreducible module $L(M_1\La_1+M_2\La_2)$
$$
h_V(M_1,M_2) \eql h_{\bS}(M_1,M_2) - h_{\bS}(M_1-1,M_2-1)\,,\tag{2.24}
$$
where 
$$
h_{\bS}(M_1,M_2) \eql \sum_{m_1+m_2+m_3=M_1\atop
m^1+m^2+m^3=M_2}\ e^{\sum_i (m_i-m^i) \ep_i} \,,\tag{2.25}
$$
denotes the Hilbert function of $\bS$.

% -------------------------------------------------------------------------
\subsubhead 2.2.3. $\frak{sl}_4$ \endsubsubhead \shskip

For $\frak{sl}_4$ we have three fundamental representations of
dimensions $D_1=4$, $D_2=6$ and $D_3=4$.  The relations generating
$I\subset \CC[\xd{i},\xd{ij},\xd{ijk}]$ arise from the 
tensor products
$$ \align
L(\La_1) \otimes L(\La_3) & ~\cong~ L(\La_1+\La_3) \oplus L(0) \,,\\
\sym^2 L(\La_2) & ~\cong~ L(2\La_2) \oplus L(0) \,,\\
L(\La_1) \otimes L(\La_2) & ~\cong~ L(\La_1+\La_2) \oplus L(\La_3) \,,\\
L(\La_2) \otimes L(\La_3) & ~\cong~ L(\La_2+\La_3)\oplus L(\La_1) \,,
\tag{2.26} \endalign
$$
and are given by 
$$\align
\si & \eql \ep^{ijkl} \xd{i} \xd{jkl} \,, \\
\bar\si & \eql \ep^{ijkl} \xd{ij} \xd{kl} \,, \\
\si^i & \eql \ep^{ijkl} \xd{j} \xd{kl} \,, \\
\si_i & \eql \ep^{jklm} \xd{ijk}\xd{lm}  \,. \tag{2.27}\endalign
$$

\centerline{\epsffile{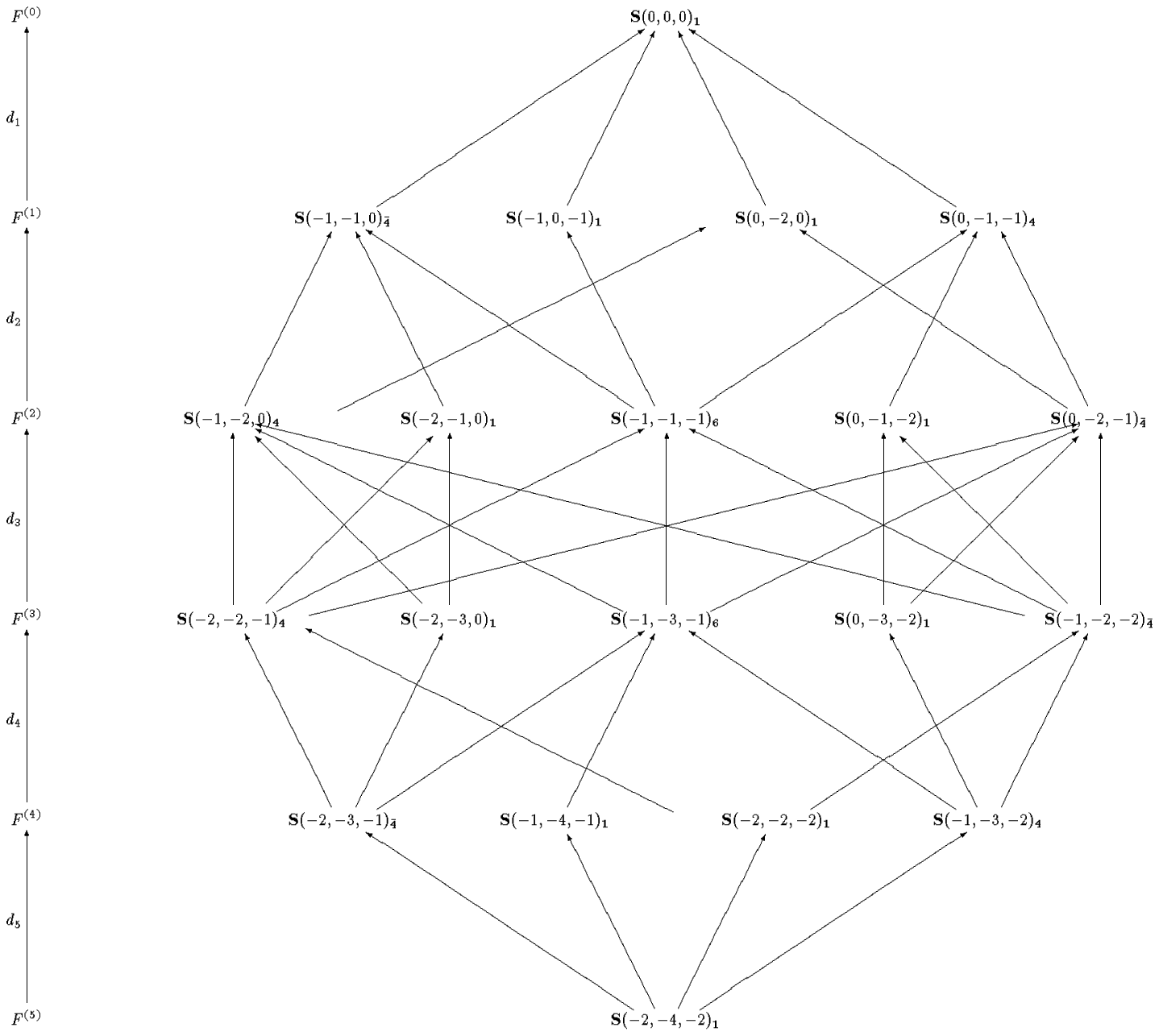}}
\bigskip
\centerline{Fig.\ 2.2.\ Resolution of $\bS(V)$ for $\frak{sl}_4$.}
\bigskip

The resolution of
$\bS(V) =\bS/I$ is depicted in Fig.\ 2.2.  We refrain from giving 
the explicit maps which are easily worked out from the shifts in
degrees indicated in the figure.  Also, we have used the 
notation $\bS(a,b)_{V}\eql \bS(a,b)\otimes V$ for an $\frak{sl}_4$ module 
$V$.  Note that the resolution in is complete agreement with Conjecture 2.1.

% -------------------------------------------------------------------------
\subhead 2.3. Example: $\frak{so}_{2n+1}$ \endsubhead\shskip

Consider the Lie algebra $B_n \cong \frak{so}_{2n+1}$ of rank $\ell=n$.
Let $\{\ep_i\}_{i=1}^n$ be an orthonormal basis of $\CC^n$.  The simple roots 
and fundamental weights of $\frak{so}_{2n+1}$ can be written as 
$$ \align
\al_i \eql \ep_i-\ep_{i+1}\,, &\qquad \La_i \eql \ep_1+\ldots+\ep_i\,,\qquad
i=1,\ldots,n-1\,,\\
\al_n \eql \ep_n \,, & \qquad \La_n \eql {\textstyle{1\over2}}
 (\ep_1+\ldots+\ep_n)\,.
\tag{2.28} \endalign
$$
The irreducible finite dimensional representations $L(\La_i)$ of
$\frak{so}_{2n+1}$ have dimension
$$ \align
D_i & \eql \binom{2n+1}{i}\,,\qquad i=1,\ldots,n-1\,,\\
D_n & \eql \qquad 2^n\,.\tag{2.29} \endalign
$$
For simplicity, we only consider here the representations $L(\La_1)$ 
and $L(\La_n)$, i.e., the vector and the spinor representation, 
respectively.  The weights of $L(\La_1)$ are 
$$ 
\{ \ep_1,\ep_2,\ldots,\ep_n,0,-\ep_n,\ldots,-\ep_1\} \,,\tag{2.30}
$$
with corresponding coordinates $\{x_i\} = \{x_1,\ldots,x_n,x_0,x_{\bar{n}},
\ldots,x_{\bone}\}$, while the weights of $L(\La_n)$ are 
$$
\{ {\textstyle{1\over2}}( \pm\ep_1\pm\ep_2\pm\ldots\pm\ep_n)\} \,,\tag{2.31}
$$
with corresponding coordinates $\{x_\al\} = \{x_{\pm\pm\ldots\pm}\}$.
The explicit realization of $\frak{so}_{2n+1}$ on $\symb L(\la_1) 
\otimes \symb L(\La_n) \cong \CC[x_i,x_\al]$ is easily 
constructed (see section 2.3.1 for $\frak{so}_5$).  The relevant 
tensor products are
$$\align
\sym^2 L(\La_1)  & ~\cong~ L(2\La_1) \oplus L(0) \,,\\
L(\La_1)\otimes L(\La_n) & ~\cong~ L(\La_1+\La_n) \oplus L(\La_n) \,, \\
\sym^2 L(\La_n)  & ~\cong~ L(2\La_n) \oplus \big(
 \bigoplus_{k=1\atop \text{restrictions}}^{n-1} L(\La_k) \big) \,,
\tag{2.32}\endalign 
$$
where, in the last tensor product, the sum is over those $k$ such that 
$k\equiv n\, \text{mod}\,4$ or $k\equiv\, (n+1)\,\text{mod}\,4$.
The first two of these lead to quadratic relations of the form
$$
\si \eql g^{ij} x_ix_j\,,\qquad \si_\al\eql (\ga^i)_\al{}^\be x_i x_\be\,,
\tag{2.33}
$$
for some matrix $g^{ij}$ and some set of $\ga$-matrices $\ga^i$,
while the last leads to relations depending on the specific value of $n$.
The realization of $\frak{so}_{2n+1}$ can be chosen in such a way that 
the matrices satisfy the properties 
$$
\{\ga^i,\ga^j\} \eql 2g^{ij} \eql \de^{i\bar\jmath}\,,
\qquad (\ga^i)_\al{}^\be \eql 
(\ga^{\bar\imath})_\be{}^\al\,.\tag{2.34}
$$
In addition, it is possible to define a charge conjugation matrix $C^{\al\be}$,
so that if spinor indices are raised and lowered by means of $C^{\al\be}$
and its inverse, e.g., $x^\al=C^{\al\be}x_\be$, then we have the additional
symmetry property
$$
(\ga^i)^{\al\be} \eql - (\ga^i)^{\be\al} \,.\tag{2.35}
$$
In the next subsection we discuss the case $\frak{so}_5$ in somewhat more
detail.

% -------------------------------------------------------------------------
\subsubhead 2.3.1. $\frak{so}_5$ \endsubsubhead \shskip

Consider the Lie algebra $\frak{so}_5$.   As discussed, we
introduce variables $\{x_i\}=
\{\xd{1},\xd{2},\xd{0},\xd{\btwo},\xd{\bone}\}$ and  $\{x_\al\}=
\{\xd{++},\xd{+-},\xd{-+},\xd{--}\}$ corresponding to the 
two fundamental representations 
$L(\La_i)$, $i=1,2$, of dimensions $D_1=5$ and $D_2=4$. \pars

The realization for the Chevalley generators
of $\frak{so}_5$ on $\bS \equiv \CC[x_i,x_\al]$ 
can be chosen as
$$\align
e_1 & \eql \xd{1}\dxd{2} - \xd{\btwo} \dxd{\bone} - \xd{+-}\dxd{-+} \,,\\
e_2 & \eql \sqrt2 \left( \xd{2}\dxd{0} - \xd{0} \dxd{\btwo}\right) - \left(
  \xd{++}\dxd{+-} - \xd{-+}\dxd{--} \right) \,,\\
h_1 & \eql \xd{1}\dxd{1} - \xd{\bone}\dxd{\bone} - \xd{2}\dxd{2} 
  + \xd{\btwo}\dxd{\btwo} + \xd{+-}\dxd{+-} - \xd{-+} \dxd{-+} \,,\\
h_2 & \eql  2\left( \xd{2}\dxd{2}  -  \xd{\btwo}\dxd{\btwo} \right) + 
  \left( \xd{++}\dxd{++} + \xd{-+}\dxd{-+} -\xd{+-}\dxd{+-} -
  \xd{--}\dxd{--} \right)  \,,\\
f_1 & \eql \xd{2}\dxd{1} - \xd{\bone} \dxd{\btwo} - \xd{-+}\dxd{+-} \,,\\
f_2 & \eql \sqrt2\left( \xd{0}\dxd{2} - \xd{\btwo} \dxd{0}\right) - \left(
  \xd{+-}\dxd{++} - \xd{--}\dxd{-+} \right) \,. \tag{2.36} \endalign
$$
In this realization one finds the following explicit expressions for
the metric $g^{ij}$  (cf.\ (2.33))
$$
g^{ij} \eql \half \de^{i \bar{\jmath}}\,,\tag{2.37}
$$
the $\ga$-matrices $(\ga^i)_\al{}^\be$
$$ \align
\ga^1  & \eql  \ta^+ \otimes \ta^3 \,, \qquad 
\ga^{\bone}  \eql  \ta^- \otimes \ta^3 \,,\\
\ga^2  & \eql  \ta^+ \otimes \ta^+ \,, \qquad
\ga^{\btwo}  \eql \ta^- \otimes \ta^- \,,\\
\ga^0 & \eql  \sqrt\half  \ta^3 \otimes \ta^3\,, \tag{2.38}\endalign
$$
and the charge conjugation matrix $C^{\al\be}$
$$
C \eql \ta^1 \otimes i\ta^2\,,\tag{2.39}
$$
where $\ta^\pm = \half(\ta^1\pm i\ta^2)$, and the $\ta^i$, $i=1,2,3$,
are the standard Pauli matrices acting on the two-component coordinates
$\xd{\pm\pm}$ (see also section 4.4 for more explicit expressions).\pars

Note that in the case of $\frak{so}_5$, the third tensor product 
in (2.32) does not produce any additional relations, so we conclude 
that the ideal of quadratic relations, defining the homogeneous 
coordinate ring $\bS(V)$ of the (complete) flag variety for $\frak{so}_5$,
is generated by $\si$ and $\si_\al$ of (2.33). \pars

Now consider the following sequence of homomorphisms (see also Fig.\ 2.3)
$$ 
0 @>>> F^{(3)} @>d_3>> F^{(2)} @>d_2>> F^{(1)} @>d_1>> 
F^{(0)} @>>> \bS/I @>>> 0\,, \tag{2.40}
$$
where
$$ \align
F^{(0)} & ~\cong~ \bS(0,0)_{\text{\bf 1}} \,,\\
F^{(1)} & ~\cong~ \bS(-1,-1)_{\text{\bf 4}} \oplus
                   \bS(-2,0)_{\text{\bf 1}}\,, \\
F^{(2)} & ~\cong~ \bS(-1,-2)_{\text{\bf 1}} \oplus
                   \bS(-2,-1)_{\text{\bf 4}}\,, \\
F^{(3)} & ~\cong~ \bS(-3,-2)_{\text{\bf 1}} \,,\tag{2.41} \endalign
$$
The homomorphisms $d_i$ are defined by 
$$ \align
d_1\ : \ F^{(1)} ~\to~ F^{(0)} \ : \ & \qquad  
  e^{(1)}_\al ~\mapsto~ \si_\al \,e^{(0)} \,,\\
& \qquad e^{(1)} ~\mapsto~ -\si \,e^{(0)} \,,\\
d_2\ : \ F^{(2)} ~\to~ F^{(1)} \ : \ & \qquad 
 e^{(2)} ~\mapsto~ x^\al \,e^{(1)}_\al \,,\\
& \qquad e^{(2)}_\al ~\mapsto~ x_\al \,e^{(1)} -
 (\ga^i)_\al{}^\be x_i \,e^{(1)}_\be \,,\\
d_3\ : \ F^{(3)} ~\to~ F^{(2)} \ : \ & \qquad 
 e^{(3)}  ~\mapsto~ \si \,e^{(2)} +
  \si^\al \,e^{(2)}_\al \,,\tag{2.42}
\endalign
$$
where $\si$ and $\si_\al$ are given in (2.33), and
where $e^{(i)}_a$ denotes the generator of an $\bS(a,b)_V$ 
component of $F^{(i)}$. \pars

Using the properties (2.34) and (2.35) it is straightforward to verify that
(2.40) defines a complex of $\frak{so}_5$ modules.  In fact, it is not
too hard to show that it actually provides a resolution of $\bS(V)=\bS/I$.

\bigskip
\centerline{\epsffile{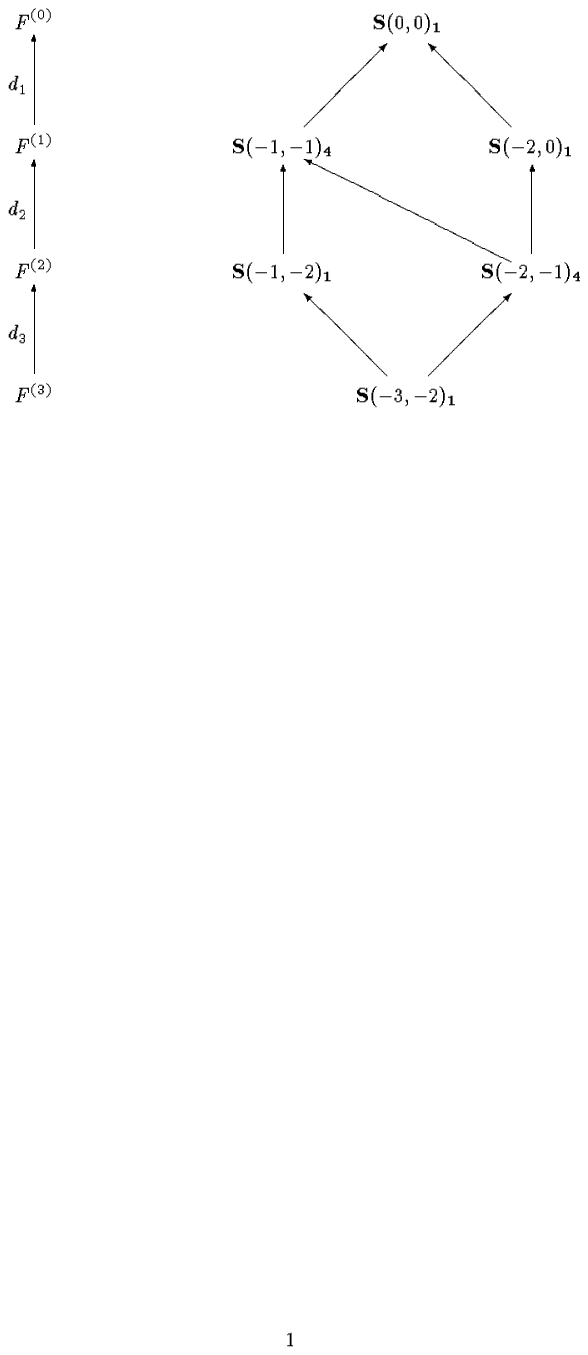}}
\bigskip
\centerline{Fig.\ 2.3.\ Resolution of $\bS(V)$ for $\frak{so}_5$}
\bigskip

Applying the Euler-Poincar\'e principle to the resolution (2.40) we find 
an explicit formula for the character $h_V(M_1,M_2)$ of $L(M_1\La_1+
M_2\La_2)$, in particular, with (2.3),
$$ \align
\text{dim}\,L(M_1\La_1+M_2\La_2) \eql & \binom{M_1+4}{4}\binom{M_2+3}{3}
 - 4 \binom{M_1+3}{4}\binom{M_2+2}{3} \\ &- \binom{M_1+2}{4}\binom{M_2+3}{3}
 + \binom{M_1+3}{4}\binom{M_2+1}{3} \\& + 4 \binom{M_1+2}{4}\binom{M_2+2}{3} 
 - \binom{M_1+1}{4}\binom{M_2+1}{3} \,.
\tag{2.43} \endalign
$$

% -------------------------------------------------------------------------
\head 3. Affinized flag varieties and $q$-identities \endhead

In \cite{Bo} we introduced the concept of an affinized projective variety.
Consider a projective variety $V$, defined by an ideal $I = \bI(V)
\subset \CC[x_1,\ldots,x_n]$,
generated by a set of homogeneous elements $f_i$, $i=1,\ldots,t$.
The affinized projective variety $\wh V$ is the infinite dimensional 
projective variety defined by the ideal $\wh I =\bI(\wh V)$ generated 
by the relations $f_i[m]$, $i=1,\ldots,t$, $m\in\ZZ_{\geq0}$, in
$\wh \bS = \CC[x_1[m],\ldots,x_n[m]]_{m\in\ZZ_{\geq0}}$, where the 
generators $f_i[m]$ of $\wh I$ are obtained from $f_i$ by replacing
all monomials $x_{i_1}\ldots x_{i_r}$ in $f_i$ by
$$
(x_{i_1}\ldots x_{i_r})[m] \eql \sum_{n_{i_1},\ldots,n_{i_r} \geq0 \atop
n_{i_1}+\ldots+n_{i_r}=m}\ x_{i_1}[n_1]\ldots x_{i_r}[n_{i_r}]\,. \tag{3.1}
$$
The homogeneous coordinate ring $\bS(\wh V)$ of the affinized projective
variety $\wh V$ is graded both by the multi-degree inherited from 
the underlying projective variety $V$ as wel as the energy $m$, i.e.,
$$
\deg(x_i[m]) \eql (\deg(x_i);m)\,.\tag{3.2}
$$
In the context of flag varieties, we have an additional grading corresponding 
to the $\frak g$-weight $\la$.  Denoting by $\bS(\wh V)_{(\bM;N;\la)}$
the vector space of homogeneous polynomials $f$ of multi-degree $(\bM;N)$
and weight $\la$ in $\bS(\wh V)$, the (partial) Hilbert series of $\wh V$
is defined as the (formal) $\frak g$-character
$$
h_{\wh V}(\bM;q) \eql \sum_{N\geq 0\,, \la} \ \text{dim}\,
\bS(\wh V)_{(\bM;N;\la)}\ q^N\,e^\la\,. \tag{3.3}
$$

The main result of \cite{Bo} was the explicit computation of the
(partial) Hilbert series $h_{\wh V}(\bM;q)$ for the coordinate ring of
an affinized projective variety associated to a quadratic monomial
ideal $I\subset\bS \equiv\CC[x_1,\ldots,x_n]$.  
If $I = \langle x_ix_j\rangle _{(i,j)\in \cP}
\subset \bS$ for some set $\cP$ of (ordered) pairs $(i,j), i<j$, with 
$i,j\in\{1,\ldots,n\}$, and $\bM=(M_1,\ldots,M_n)$ is the multi-degree
where $M_i$ denotes the number of $x_i$ in a monomial, it was found that
(see \cite{Bo}, section 4.1)
$$
h_{\wh V}(\bM;q) \eql {q^{\sum_{(i,j)\in\cP} M_iM_j} \over \qn{M_1}\ldots
\qn{M_n} } \,. \tag{3.4}
$$
In \cite{Bo} we also gave an
algorithm to compute an alternating sum expression for $h_{\wh V}(\bM;q)$ 
based on Taylor's resolution of $\bS/I$.  This algorithm is based on
the identity
$$
{q^{M_1M_2}\over \qn{M_1}\qn{M_2} } \eql \sum_{m\geq0} \ (-1)^m
{q^{{1\over2}m(m-1)} \over \qn{m}\qn{M_1-m}\qn{M_2-m} }\,,\tag{3.5}
$$
discovered and proved in \cite{BS1,BS2} and shown to be related 
to the coordinate ring of the affinized projective variety
associated to the ideal $I = \langle x_1x_2\rangle \subset \CC[x_1,x_2]$
in \cite{Bo}. 

\remark{Remark 3.1}
The identity (3.5), as well as its `inverse' (see \cite{BS2}) 
$$
{1\over \qn{M_1}\qn{M_2} } \eql \sum_{m\geq0} \ 
{q^{(M_1-m)(M_2-m)} \over \qn{m}\qn{M_1-m}\qn{M_2-m} }\,,\tag{3.6}
$$
are intimately related to, in fact can be used to prove, the five-term
identity for Rogers' dilogarithm \cite{BCR}.
\endremark\medskip

In the remainder of this section we will explain how the results
of \cite{Bo}, summarized above, can be used to compute the Hilbert 
series of affinized flag varieties.

As we have seen in section 2, the ideal that defines flag varieties, 
albeit quadratic, is not a monomial ideal.  However,
for the purpose of calculating the Hilbert series, we can replace
$I$ by the ideal of leading terms $\langle\text{LT}(I)\rangle$  
with respect to any ordering on $\bS$.    
Indeed, (cf.\ Ch.\ 9.3, Prop.\ 9 in \cite{CLO1})
$$
h_{\bS/I}(\bM;q) \eql  h_{\bS/\langle\text{LT}(I)\rangle}(\bM;q) \,. 
\tag{3.7}
$$
We recall that 
$\langle\text{LT}(I)\rangle$ is the ideal generated
by the leading terms $\text{LT}(f)$ for all $f\in I$.
The ideal $\langle\text{LT}(I)\rangle$ is finitely generated (by Hilbert's 
basis theorem) and a set of generators is given by $\text{LT}(g_i)$,
$i=1,\ldots,s$, where $\{ g_i\}_{i=1}^s$ is a Gr\"obnerbasis for $I$
(see, e.g., \cite{CLO1,CLO2}).

Unfortunately, by computing the Hilbert 
series by means of $\langle\text{LT}(I)\rangle$ we give up the 
manifest $\bfg$ symmetry, since $\bS/\langle\text{LT}(I)\rangle$ is no
longer a $\bfg$-module.  In examples we will see that the explicit 
$\frak g$-structure of the Hilbert series can be restored by 
successive application of (3.5).  

\remark{Remark 3.2} In fact, note that the passing from $I$ to 
$\langle\text{LT}(I)\rangle$ is reminiscent of taking the crystal
limit of $U_q(\bfg)$ modules.  It would be interesting to explore this
connection further. \endremark\medskip

More seriously, while $\langle\text{LT}(I)\rangle$ is a 
monomial ideal, it is in general no longer quadratic (with the exception of 
$\frak{sl}_3$, see section 4.2).   However, 
non-quadratic monomial ideals may be transformed into quadratic monomial
ideals by introducing additional variables.  Having achieved this,
we can then apply the results of \cite{Bo} and derive an explicit 
expression for the Hilbert series $h_{\wh V}(\bM;q)$ of the 
affinized flag varieties $\wh V$.\pars

Let us illustrate in an example how the introduction of additional 
variables solves the problem above.
Consider $I = \langle x_1x_2\ldots x_n\rangle \subset
\CC[x_1,\ldots,x_n]$. Clearly, for $n\geq3$,
$$
\CC[x_1,\ldots,x_n]/I ~\cong~ \CC[x_1,\ldots,x_n,t_1,
\ldots,t_{n-2}]/\wt{I}\,, \tag{3.8}
$$
where 
$$
\wt{I} \eql \langle t_1-x_1x_2, t_2-t_1x_3, \ldots, t_{n-2} - t_{n-3}x_{n-1},
t_{n-2}x_n \rangle\,. \tag{3.9}
$$
The isomorphism (3.8) is multi-degree preserving provided we assign
$$
\deg(t_i) \eql (\underbrace{1,1,\ldots,1}_{i+1},0,\ldots,0)\,.\tag{3.10}
$$
With respect to the lexicographic ordering defined by 
$$
x_1>x_2>\ldots>x_n>t_1>\ldots>t_{n-2}\,,\tag{3.11}
$$
we find 
$$
\langle\text{LT}(\wt{I})\rangle \eql
\langle x_1x_2, t_1x_3,\ldots, t_{n-2}x_n\rangle\,,\tag{3.12}
$$
which happens to be a quadratic monomial ideal.  Thus, we can apply 
the results from \cite{Bo} and conclude that 
for the corresponding affinized variety $\wh V$ we have
$$
h_{\wh V}(\bM;q) \eql 
% h_{\wh V/ \langle \text{LT}(\wt{I})\rangle}(\bM;q) \eql
\sum_{m_1,\ldots,m_{n-2}\geq0}\  
{q^{Q} \over \prod_{i=1}^n \qn{M_i -\De M_i} }\,, \tag{3.13}
$$
where the quadratic form $Q$ is given by
$$ \align
Q \eql & (M_1-(m_1+\ldots+m_{n-2}))(M_2-(m_1+\ldots+m_{n-2})) \\ & + 
  m_1(M_3-(m_2+\ldots +m_{n-2})) + \ldots +m_{n-2}M_n\,, \tag{3.14}\endalign
$$
and
$$ \align
\De M_1 & \eql m_1+\ldots+m_{n-2}\,, \\
\De M_2 & \eql m_1+\ldots+m_{n-2}\,, \\
\De M_3 & \eql m_2+\ldots +m_{n-2}\,, \\
\vdots\quad  & \eql  \quad\vdots  \\
\De M_{n-1}& \eql m_{n-2} \,, \\
\De M_n & \eql 0 \,. \tag{3.15} \endalign
$$
Note, moreover, that by repeated application of 
(3.5) and (3.6), eqn.\ (3.13) can be written as 
$$
h_{\wh V}(M;q) \eql \sum_{m\geq0}\ (-1)^m {q^{{1\over2}m(m-1)}\over \qn{m}}
{1\over \qn{M_1-m} \ldots \qn{M_n-m} } \,.\tag{3.16}
$$
This identity turns out to play an important role in the spinon 
description of
$(\wh{\frak{sl}_n})_{k=1}$ modules \cite{BS1,BS2,BCR}.\medskip

The above example illustrates that, 
for the purpose of computing the Hilbert series,
non-quadratic monomial ideals can always be reduced to quadratic
monomial ideals by the introduction of additional variables.  For
further examples we refer to section 4.

% -------------------------------------------------------------------------
\head 4. Examples \endhead

In this section we will explicitly compute the (partial) Hilbert 
series for the affinized flag varieties of $\frak{sl}_n$, $n=2,3,4$,
and $\frak{so}_{2n+1}$, $n=2,3$.

% -------------------------------------------------------------------------
\subhead 4.1. $\frak{sl}_2$ \endsubhead\shskip

Recall from section 2.2.1 that 
the coordinate ring of the flag variety $V$ for $\frak{sl}_2$ is
isomorphic to $\bS = \CC[x_1,x_2]$, i.e., the ideal of quadratic
relations is trivial.  Hence, the Hilbert series of the affinized
variety $\wh V$ is simply given by
$$
h_{\wh V}(M;q) \eql \sum_{m_1+m_2=M}\ {1\over \qn{m_1}\qn{m_2}}
  e^{m_1\ep_1+m_2\ep_2} \,, \tag{4.1}
$$
which can obviously be interpreted as an `affinization' of equation 
(2.19).

% -------------------------------------------------------------------------
\subhead 4.2. $\frak{sl}_3$ \endsubhead\shskip

The coordinate ring of the flag variety $V$ for $\frak{sl}_3$ is given
by $\bS(V) = \bS/I$ with $\bS=\CC[x_i,x_{\bar\imath}]_{i=1,2,3}$, and
$I=\langle x_1x_{\bar1}+x_2x_{\bar2}+x_3x_{\bar3}\rangle$
(cf.\ section 2.2.2 where we used the notation $x^i=x_{\bar\imath}$).  
With respect to the lexicographic ordering defined by
$$
x_1>x_2>x_3>x_{\bar1}>x_{\bar2}>x_{\bar3}\,, \tag{4.2}
$$
we obviously have 
$$
\langle \text{LT}(I)\rangle \eql \langle x_1x_{\bar1} \rangle\,. \tag{4.3}
$$
Thus, we can immediately apply our result for quadratic monomial 
ideals and conclude that the partial Hilbert series of the affinization 
$\wh V$ is given by
$$
h_{\wh V}(M_1,M_2;q) \eql \sum_{m_1+m_2+m_3=M_1\atop
m_{\bar1}+m_{\bar2}+m_{\bar3}=M_2} \ {q^{m_1m_{\bar1}} \over
\prod_i \qn{m_i}  \qn{m_{\bar\imath}}} 
e^{\sum_i (m_i-m_{\bar\imath}) \ep_i}\,. \tag{4.4}
$$
As discussed in section 3, in
order to use our results for quadratic monomial ideals, we have had
to pass from $I$ to $\langle \text{LT}(I)\rangle$, thereby breaking the 
manifest $\frak{sl}_3$ symmetry.  The result (4.4), however, can be 
written as an explicit $\frak{sl}_3$ character by using (3.5), i.e.,
$$
h_{\wh V}(M_1,M_2;q) \eql \sum_{m_1+m_2+m_3+m=M_1\atop
m_{\bar1}+m_{\bar2}+m_{\bar3}+m=M_2} \ (-1)^m {q^{{1\over2} m(m-1)}\over
\qn{m} } {1 \over \prod_i \qn{m_i}  \qn{m_{\bar\imath}}} 
e^{\sum_i (m_i-m_{\bar\imath}) \ep_i}\,, \tag{4.5}
$$
or, as
$$
h_{\wh V}(M_1,M_2;q) \eql \sum_{m\geq0} \ (-1)^m {q^{{1\over2} m(m-1)}\over
\qn{m} } h_{\wh\bS}(M_1-m,M_2-m;q) \,,\tag{4.6}
$$
where 
$$
h_{\wh\bS}(M_1,M_2;q) \eql \sum_{m_1+m_2+m_3=M_1\atop
m_{\bar1}+m_{\bar2}+m_{\bar3}=M_2}\ {1\over \prod_i \qn{m_i}  
\qn{m_{\bar\imath}}} e^{\sum_i (m_i-m_{\bar\imath}) \ep_i} \,,\tag{4.7}
$$
denotes the Hilbert series of the affinization of the coordinate ring 
$\bS = \CC[x_i,x_{\bar\imath}]$.
Clearly, (4.7) has to be interpretated as the Hilbert series obtained 
by applying the Euler-Poincar\'e principle to an affinization of the 
resolution (2.22).  In this case the affinization of the resolution is easily 
constructed and yields a Koszul complex (cf.\ \cite{Bo}).

% -------------------------------------------------------------------------
\subhead 4.3. $\frak{sl}_4$ \endsubhead\shskip

The coordinate ring for the flag variety of $\frak{sl}_4$ is 
given by $I\subset \CC[\xd{i},\xd{ij},\xd{ijk}] = \bS$ where the generators 
of $I$ are explicitly given in section 2.2.3.
With respect to the lexicographic ordering on $\bS$ defined by
$$ \align
& \xd{1}>\xd{2}>\xd{3}>\xd{4}> \\
& \xd{12}>\xd{13}>\xd{14}>\xd{23}>\xd{24}>\xd{34}> \\
& \xd{123}>\xd{124}>\xd{134}>\xd{234} \,, \tag{4.8} \endalign
$$
we find 
$$ \align
\langle \text{LT}(I) \rangle \eql \langle &
\xd{1} \xd{23}, \xd{1} \xd{24}, \xd{1} \xd{34}, \xd{1} \xd{234}, 
\xd{2}\xd{34},  \xd{2} \xd{13} \xd{24}, 
\xd{12} \xd{34}, \\ & \xd{12} \xd{134}, \xd{12} \xd{234}, 
\xd{13} \xd{234}, 
\xd{13} \xd{24}\xd{134}, \xd{14} \xd{234}  \rangle \,. \tag{4.9} 
\endalign
$$
Note that in this case the ideal of leading terms 
$\langle \text{LT}(I) \rangle$ is not a quadratic monomial ideal.
We can however apply the trick discussed in section 3 and introduce 
and additional variable
$$
t \eql \xd{13} \xd{24} \,, \tag{4.10}
$$
of multi-degree $\deg(t) \eql (0,2,0)$
and vanishing $\frak{sl}_4$-weight.  Indeed,
$$
\CC[\xd{i},\xd{ij},\xd{ijk}] / \langle \text{LT} (I)\rangle  ~\cong~
\CC[\xd{i},\xd{ij},\xd{ijk},t] / \wt I \,,\tag{4.11}
$$
where 
$$\align
\wt I \eql \langle &
\xd{1} \xd{23}, \xd{1} \xd{24}, \xd{1} \xd{34}, \xd{1} \xd{234}, 
\xd{2}\xd{34},  \xd{2} t , 
\xd{12} \xd{34}, \\ & \xd{12} \xd{134}, \xd{12} \xd{234}, 
\xd{13} \xd{234}, 
t\xd{134}, \xd{14} \xd{234},   t - \xd{13} \xd{24} \rangle \,. \tag{4.12}
\endalign
$$
Extending the lexicographic ordering on $\bS$ defined by (4.8) to
$\wt \bS = \CC[\xd{i},\xd{ij},\xd{ijk},t]$ by
$$
\xd{1} > \ldots > \xd{234} > t\,, \tag{4.13}
$$ 
we can now pass to the ideal of leading terms of $\wt I$.  We find 
$$ \align
\langle \text{LT}(\wt I) \rangle \eql \langle & 
\xd{1} \xd{23}, \xd{1} \xd{24}, \xd{1} \xd{34}, \xd{1} \xd{234}, \xd{1} t,
\xd{2}\xd{34},  \xd{2} t , 
\xd{12} \xd{34}, \\ & \xd{12} \xd{134}, \xd{12} \xd{234}, 
\xd{13} \xd{24}, \xd{13} \xd{234}, \xd{14} \xd{234},
\xd{134} t , \xd{234} t \rangle \,. \tag{4.14} \endalign
$$
Thus, $\langle \text{LT}(\wt I) \rangle$ is a quadratic monomial ideal,
and we are finally in a position to apply the results of \cite{Bo}.
We find 
$$ \align
h_{\wh V}(M_1,M_2,M_3;q) \eql 
\sum_{ \sum m_i=M_1\atop { \sum m_{ij}+2m =M_2 \atop
 \sum m_{ijk} = M_3 }} &
{q^Q \over \qn{m}\prod_i \qn{m_i} \prod_{i<j} \qn{m_{ij}} \prod_{i<j<k}
 \qn{m_{ijk}} } \\
& \times e^{\sum m_i\ep_i + \sum 
  m_{ij}(\ep_i+\ep_j) + \sum m_{ijk}(\ep_i+\ep_j+\ep_k)}
\,, \tag{4.15} \endalign
$$
with quadratic form 
$$\align
Q \eql & m_1( m_{23} +m_{24}+m_{34} + m_{234} + m ) 
+ m_2 (m_{34} + m) + m_{12}m_{34} \\ & + m_{13}m_{24} + 
m_{134} ( m_{12} + m) +
m_{234} ( m_{12} + m_{13} + m_{14} + m) \,.
\tag{4.16}\endalign
$$

As in section 4.2 we may now attempt to rewrite (4.15) as a manifest 
$\frak{sl}_4$ character by successively applying (3.5)
according to the algorithm outlined in \cite{Bo}.  One would expect
that the result can be written in a form that can be interpreted as 
arising from a properly affinized version of the resolution of Fig.\ 2.2.
We have not succeeded in carrying out this program in all generality,
but partial results (relevant to the applications discussed in \cite{BS4})
are easily obtained, e.g., (cf.\ (4.6))
$$
h_{\wh V}(M_1,0,M_3;q) \eql \sum_{m\geq0} \ (-1)^m 
{q^{{1\over2} m(m-1)}\over
\qn{m} } h_{\wh\bS}(M_1-m,0,M_3-m;q) \,,\tag{4.17}
$$
where 
$$
h_{\wh\bS}(M_1,0,M_3;q) \eql \sum_{\sum m_i=M_1\atop
\sum m_{ijk}=M_3}\ {1\over \prod_i \qn{m_i} \prod_{i<j<k} 
\qn{m_{ijk}}} e^{\sum_i m_i\ep_i +
\sum m_{ijk} \ep_{ijk} } \,,\tag{4.18}
$$
denotes the Hilbert series of the affinization of the coordinate ring 
$\bS = \CC[x_i,x_{ijk}]$.
The result (4.18) generalizes in an obvious way to $\frak{sl}_n$.

% -------------------------------------------------------------------------
\subhead 4.4. $\frak{so}_5$ \endsubhead\shskip

For $\frak{so}_5$ we need to consider $I\subset \CC[x_i,x_\al]=\bS$ where
$I = \langle \si, \si_\al \rangle$ is given in section 2.3.1.  Explicitly,
$$\align
\si      & \eql \xd{1}\xd{\bone} + \xd{2}\xd{\btwo} + \half\xd{0}\xd{0} \,,\\
\si_{++} & \eql \xd{1}\xd{-+}+\sqrt\half\xd{0}\xd{++} +\xd{2} \xd{+-}\,,\\
\si_{+-} & \eql -\xd{1}\xd{--}-\sqrt\half\xd{0}\xd{+-}+\xd{\btwo}\xd{++}\,,\\
\si_{-+} & \eql \xd{2}\xd{--}-\sqrt\half\xd{0}\xd{-+}+\xd{\bone}\xd{++}\,,\\
\si_{--} & \eql \xd{\btwo}\xd{-+}+\sqrt\half\xd{0}\xd{--}-
  \xd{\bone}\xd{+-}\,. \tag{4.19} \endalign
$$
The (minimal) Gr\"obnerbasis with respect to the lexicographic ordering
on $\bS$ defined by
$$
\xd{1}>\xd{2}>\xd{0}>\xd{\btwo}>\xd{\bone}>\xd{++}>\xd{+-}>\xd{-+}>\xd{--}\,,
\tag{4.20}
$$
is given by $\{ \si, \si_\al, \ta\}$ where
$$ \align
\ta & \eql - \sqrt\half \xd{0}\xd{\bone}\xd{++} - \xd{2}\xd{\bone}\xd{+-}
  + \half \xd{0}\xd{0} \xd{-+} + \xd{2}\xd{\btwo}\xd{-+}\\
& \eql \si \xd{-+} - \xd{\bone} \si_{++}  \,. \tag{4.21}\endalign
$$
The monomial ideal $\langle \text{LT} (I)\rangle$ generated by the 
leading terms in $I$ is therefore given by
$$ 
\langle \text{LT} (I)\rangle  \eql \langle \xd{1}\xd{\bone},
\xd{1}\xd{-+}, \xd{1}\xd{--}, \xd{2}\xd{\btwo}\xd{-+} ,
\xd{2}\xd{--}, \xd{0}\xd{--}\rangle \,. \tag{4.22}
$$
Again, to apply the results of \cite{Bo}, we need to introduce
one additional variable $t=\xd{2}\xd{\btwo}$, of multi-degree
$\deg(t) = (2,0)$ and vanishing $\frak{so}_5$ weight.  Then
$$
\CC[\xd{i},\xd{\al}] / \langle \text{LT} (I)\rangle  ~\cong~
\CC[\xd{i},\xd{\al},t] / \wt I \,,\tag{4.23}
$$
with 
$$
\wt I \eql \langle \xd{1}\xd{\bone},
\xd{1}\xd{-+}, \xd{1}\xd{--}, \xd{2}\xd{--}, \xd{0}\xd{--}, 
t- \xd{2}\xd{\btwo}, t \xd{-+} \rangle\,. \tag{4.24}
$$
Extending the lexicographic ordering on $\bS$ defined by (4.20) to
$\wt \bS = \CC[x_i,x_\al,t]$ by
$$
\xd{1}>\xd{2}>\xd{0}>\xd{\btwo}>\xd{\bone}>\xd{++}>\xd{+-}>\xd{-+}>
\xd{--}>t\,,\tag{4.25}
$$
we obtain
$$
\langle \text{LT} (\wt{I})\rangle \eql
\langle \xd{1}\xd{\bone},
\xd{1}\xd{-+}, \xd{1}\xd{--}, \xd{2}\xd{\btwo}, \xd{2}\xd{--}, \xd{0}\xd{--}, 
\xd{-+}t, \xd{--}t \rangle \,,\tag{4.26}
$$
which is a quadratic monomial ideal.  Thus, we conclude that the Hilbert 
series of the coordinate ring of the affinized flag variety of $\frak{so}_5$
is given by
$$
h_{\wh V}(M_1,M_2;q) \eql 
\sum_{ \sum m_i + 2m = M_1 \atop \sum m_\al = M_2}\ 
 {q^Q\over \prod_i \qn{m_i} \prod_\al \qn{m_\al} \qn{m} }
 \ e^{\sum m_i\la_i + \sum m_\al \la_\al} \,,\tag{4.27}
$$
where 
$$
Q \eql m_1( m_{\bone} + m_{-+} + m_{--} ) + m_2 (m_{\btwo} + m_{--}) 
  + m_0 m_{--} + m ( m_{-+} + m_{--}) \,,\tag{4.28}
$$
and $\{\la_i\}$ and $\{\la_\al\}$ denote the weights of the 
representations $L(\La_1)$ and $L(\La_2)$, respectively (see section 2.3).

Again, in principle one could proceed to write (4.27) as a manifest
$\frak{so}_5$ character by successive application of (3.5).  We
have not been able to carry this out in all generality, but the
following partial results can be proved 
$$ \align 
h_{\wh V}(0,M_2;q) \eql & h_{\wh\bS}(0,M_2;q) \,,\\ 
h_{\wh V}(M_1,0;q) \eql & \sum_{m\geq0}\ (-1)^m\, {q^{ {1\over2}m(m-1)} \over
  \qn{m}} \, h_{\wh\bS}(M_1-2m,0;q)\,,\\ 
h_{\wh V}(M_1,1;q) \eql &  \phantom{-}
  \sum_{m\geq0}\ (-1)^m\, {q^{ {1\over2}m(m-1)} \over
  \qn{m}} \, h_{\wh\bS}(M_1-2m,1;q) \\
 & -  \sum_{m\geq0}\ (-1)^m\, {q^{ {1\over2}m(m+1)} \over
  \qn{m}\qn{1}} \, h_{\wh\bS}(M_1-2m-1,0;q)\ch_{\text{\bf 4}} \\
 & + \sum_{m\geq0}\ (-1)^m\, {q^{ {1\over2}m(m+1)} \over
  \qn{m}\qn{1}} \, h_{\wh\bS}(M_1-2m-2,0;q)\ch_{\text{\bf 4}} \,,\tag{4.29}
\endalign
$$
where 
$$
h_{\wh\bS}(M_1,M_2;q) \eql 
\sum_{\sum m_i=M_1\atop \sum m_\al=M_2}\ 
{1\over \prod_i \qn{m_i} \prod_\al \qn{m_\al} } e^{\sum m_i \la_i 
+ \sum m_\al \la_\al} \,, \tag{4.30}
$$
denotes the Hilbert series of the affinization of $\bS=\CC[x_i,x_\al]$.
The third equation can be simplified, but we have left it in this 
form to elucidate its origin as arising from an `affinization' of the 
resolution (2.40) (see also Fig.\ 2.3).

% -------------------------------------------------------------------------
\subhead 4.5. $\frak{so}_7$ \endsubhead\shskip

For $\frak{so}_7$ we only quote the result for the coordinate ring
of the partial flag variety discussed in section 2.3.  The procedure is as 
before, so we will only give the main intermediate steps that lead to
the Hilbert series.  We use the notation of section 2.3.
With respect to the lexicographic ordering defined by 
$$\align
& \xd{1} > \xd{2} > \xd{3} > \xd{0} > \xd{\bthr} > \xd{\btwo} > 
\xd{\bone} > \\
& \xd{+++} > \xd{++-} \xd{+-+} >\xd{-++} > \xd{+--} > \xd{-+-} > \xd{--+}
 > \xd{---} \,, \tag{4.31}\endalign
$$
we have 
$$\align
\langle\text{LT}(I)\rangle \eql \langle &
\xd{1}\xd{\bone}, \xd{+++}\xd{---}, \xd{1}\xd{-++},\xd{1}\xd{-+-},
\xd{1}\xd{--+}, \xd{1}\xd{---}, \xd{2}\xd{--+},\\
& \xd{2}\xd{---},
\xd{3}\xd{---}, \xd{0}\xd{---}, \xd{2}\xd{\btwo}\xd{-++}, 
\xd{2}\xd{\btwo}\xd{-+-},\xd{2}\xd{+-+}\xd{-+-},\\ & 
\xd{3}\xd{\bthr}\xd{--+}, \xd{3}\xd{++-}\xd{--+}, \xd{0}\xd{++-}\xd{--+},
\xd{3}\xd{\bthr}\xd{+-+}\xd{-+-}\rangle \,. \tag{4.32} \endalign
$$
To transform this back into a quadratic monomial ideal we need to
introduce four additional variables for which one can take
$$\align
t_1 \eql \xd{++-}\xd{--+} \,,& \qquad t_2 \eql \xd{+-+}\xd{-+-}\,, \\
t_3 \eql \xd{3}\xd{\bthr}\,,& \qquad  t_4 \eql  \xd{2}\xd{\btwo}\,. \tag{4.33}
\endalign
$$
Obviously, the variables (4.33) all have vanishing $\frak{so}_7$-weight 
and multi-degrees
$$ \align
\deg(t_1) & \eql \deg(t_2) \eql (0,0,2) \,,\\
\deg(t_3) & \eql \deg(t_4) \eql (2,0,0) \,. \tag{4.34} \endalign
$$
We have 
$$
\CC[x_i,x_\al]/\langle \text{LT}(I)\rangle ~\cong~
\CC[x_i,x_\al,t_1,t_2,t_3,t_4]/ \wt I \,, \tag{4.35}
$$
where 
$$\align
\wt I \eql \langle &
\xd{1}\xd{\bone}, \xd{+++}\xd{---}, \xd{1}\xd{-++},\xd{1}\xd{-+-},
\xd{1}\xd{--+}, \xd{1}\xd{---}, \xd{2}\xd{--+},
\xd{2}\xd{---}, \\ & \xd{3}\xd{---}, \xd{0}\xd{---}, 
t_1-\xd{++-}\xd{--+}, t_2-\xd{+-+}\xd{-+-}, t_3 - \xd{3}\xd{\bthr},\\ & 
t_4 - \xd{2}\xd{\btwo}, 
t_1\xd{3},t_1\xd{0},t_2\xd{2}, t_3\xd{--+},t_2t_3,t_4\xd{-+-},
t_4\xd{--+}
\rangle \,.\tag{4.36} \endalign
$$
Extending the ordering defined by (4.31) by
$$
\xd{1} > \ldots > \xd{---} > t_1 >t_2>t_3>t_4\,,\tag{4.37}
$$
we find 
$$\align
\langle \text{LT}(\wt I)\rangle \eql \langle &
\xd{1}\xd{\bone}, \xd{1}\xd{-++},\xd{1}\xd{-+-},
\xd{1}\xd{--+}, \xd{1}\xd{---}, \xd{1}t_1,\xd{1}t_2, \\
& \xd{2}\xd{\btwo}, \xd{2}\xd{--+},\xd{2}\xd{---},\xd{2}t_1,\xd{2}t_2, \\
& \xd{3}\xd{\bthr}, \xd{3}\xd{---}, \xd{3}t_1, 
  \xd{0}\xd{---}, \xd{0}t_1,\\
& \xd{+++}\xd{---},\xd{++-}\xd{--+},\xd{+-+}\xd{-+-},\\
& \xd{-++}t_4,\xd{-+-}t_4,\xd{--+}t_3,\xd{--+}t_4,
 \xd{---}t_3,\xd{---}t_4,\\
& t_1t_3,t_1t_4,t_2t_3,t_2t_4 \rangle \,.\tag{4.38} \endalign
$$
Thus, the result for the Hilbert series is 
$$ \align
h_{\wh V}(M_1,0,M_3;q) \eql \sum_{\sum m_i + 2p_3+2p_4 = M_1 \atop
\sum m_\al + 2p_1+2p_2 = M_3 } \ 
& {q^Q \over \prod_i \qn{m_i} \prod_\al \qn{m_\al} \prod_j \qn{p_j} }\\
& \times e^{\sum m_i\la_i + \sum m_\al \la_\al} 
\,,\tag{4.39} \endalign
$$
with 
$$\align
Q \eql & \hphantom{+ }
m_1(m_{\bone} +m_{-++} +m_{-+-} +m_{--+} + m_{---} +p_1+p_2) \\
& + m_2( m_\btwo + m_{--+} + m_{---} +p_1+p_2) \\
& + m_3( m_\bthr + m_{---} +p_1 ) + m_0(m_{---} +p_1) \\
& + m_{+++}m_{---} + m_{++-}m_{--+} + m_{+-+}m_{-+-} \\
& + (m_{-++}+m_{-+-})p_4 + (m_{--+}+m_{---})(p_3+p_4) \\
& + (p_1+p_2)(p_3+p_4) \,.\tag{4.40} \endalign
$$
As in (4.29) we can write the character (4.39) as a manifest $\frak{so}_7$
character in the following cases
$$ \align
h_{\wh V}(0,0,M_3;q) \eql & \sum_{m\geq0}\ (-1)^m\,
  {q^{ {1\over2}m(m-1)} \over
  \qn{m}} \, h_{\wh\bS}(0,0,M_3-2m;q) \,,\\ 
h_{\wh V}(M_1,0,0;q) \eql & \sum_{m\geq0}\ (-1)^m\, 
  {q^{ {1\over2}m(m-1)} \over
  \qn{m}} \, h_{\wh\bS}(M_1-2m,0,0;q)\,,\\ 
h_{\wh V}(M_1,0,1;q) \eql &  \phantom{-}
  \sum_{m\geq0}\ (-1)^m\, {q^{ {1\over2}m(m-1)} \over
  \qn{m}} \, h_{\wh\bS}(M_1-2m,0,1;q) \\
 & -  \sum_{m\geq0}\ (-1)^m\, {q^{ {1\over2}m(m+1)} \over
  \qn{m}\qn{1}} \, h_{\wh\bS}(M_1-2m-1,0,0;q)\ch_{\text{\bf 8}} \\
 & + \sum_{m\geq0}\ (-1)^m\, {q^{ {1\over2}m(m+1)} \over
  \qn{m}\qn{1}} \, h_{\wh\bS}(M_1-2m-2,0,0;q)\ch_{\text{\bf 8}} \,,\tag{4.41}
\endalign
$$
where 
$$
h_{\wh\bS}(M_1,0,M_3;q) \eql 
\sum_{\sum m_i=M_1\atop \sum m_\al=M_3}\ 
{1\over \prod_i \qn{m_i} \prod_\al \qn{m_\al} } e^{\sum m_i \la_i 
+ \sum m_\al \la_\al} \,,\tag{4.42}
$$
denotes the Hilbert series of the affinization of $\bS=\CC[x_i,x_\al]$.
The second and third of these identities have an obvious generalization
to $\frak{so}_{2n+1}$, while the generalization of the first will be
more involved due to additional relations arising from the 
tensor product $\sym^2 L(\La_n)$ (see (2.32)).

% -------------------------------------------------------------------------
\head 5. Relation to modified Hall-Littlewood polynomials and 
affine Lie algebra characters\endhead

By construction, $P(\bM;q) \equiv \left(\prod_{i=1}^\ell \qn{M_i} \right)\ 
h_{\wh V}(\bM;q)$
is a $\bfg$-character valued polynomial in $q$ such that
$$
\lim_{q\to1} \ P(\bM;q) \eql
  \ch_{\La_1}^{M_1} \ldots \ch_{\La_\ell}^{M_\ell} \,. \tag{5.1}
$$
In other words, $P(\bM;q)$ is a $q$-deformation of the character of the 
tensor product $L(\La_1)^{\otimes M_1} \otimes \ldots \otimes 
L(\La_\ell)^{\otimes M_\ell}$.  A natural $q$-deformation, that shows 
up in many contexts, is the so-called modified Hall-Littlewood polynomial
(see \cite{Ki1,Ki2} for reviews and recent results).  For a general
finite dimensional simple Lie algebra $\frak g$ of rank $\ell$, it 
can be defined as follows \cite{KR}.  Let $\La_i$, $\al_i$ and $\al_i^\vee$
($i=1,\ldots,\ell$) be the fundamental weights, simple roots and 
simple co-roots, respectively.  For any pair of dominant integral 
weights $\la,\mu$, define the polynomial $M_{\la\mu}(q)$ by 
$$
M_{\la\mu}(q) \eql \sum_{\bm} q^{\wt c(\bm)} \prod_{i=1}^\ell
\prod_{a=1}^\infty \qbinom{P_a^{(i)}(\bm) + m^{(i)}_a}{m^{(i)}_a}\,,\tag{5.2}
$$
where the sum is taken over all nonnegative integers $m_a^{(i)}$
($i=1,\ldots,\ell$, $a=1,2,\ldots$), such that
$$
\mu -\la \eql \sum_{i=1}^\ell \left( \sum_{a=1}^\infty a\, m_a^{(i)} 
\right) \al_i\,.\tag{5.3}
$$ 
Moreover, 
$$
P_a^{(i)}(\bm)  \eql (\mu,\al_i^\vee) - \sum_{j=1}^\ell\,\sum_{b=1}^\infty
\Ph_{ab}^{ij}\, m_b^{(j)} \,,\tag{5.4}
$$
where 
$$
\Ph_{ab}^{ij} \eql {2(\al_i,\al_j)\over \al_i{}^2\al_j{}^2}\ 
\text{min}(a\al_i{}^2,b\al_j{}^2)\,, \tag{5.5} 
$$
and $\wt c(\bm)$ is the cocharge
$$
\wt c(\bm) \eql {\textstyle{1\over2}} \sum_{i,j=1}^\ell\,\sum_{a,b=1}^\infty
\ m_a^{(i)}\,\Ph_{ab}^{ij}\,m_b^{(j)} \,.\tag{5.6}
$$
For $\frak{sl}_n$, the polynomials (5.2) are related to the Kostka-Foulkes
polynomials $K_{\la\mu}(q)$ in a simple way (see, e.g., \cite{Ki1}).\pars

For $\la$ dominant integral, let $\chi_\la$ denote the character 
of the finite dimensional irreducible $\frak g$-module $L(\la)$.  
Then, for 
dominant integral weight $\la$,
the modified Hall-Littlewood polynomial $M_\la(q)$ is
defined as the character valued polynomial
$$
M_\la(q) \eql \sum_\mu \ M_{\mu\la}(q)\, \chi_\mu\,. \tag{5.7}
$$
It has the property that for $q=1$ it equals the character of 
the tensor product module $W_1^{M_1}\otimes \ldots \otimes 
W_\ell^{M_\ell}$, where $M_i=(\la,\al_i^\vee)$, and
where $W_i$ denotes the ``minimal affinization''
of $L(\La_i)$, i.e., the minimal irreducible module of the quantum 
affine algebra $U_q(\wh{\frak g})$ (or Yangian $Y(\frak g)$) such 
that $L(\La_i) \subset W_i$ \cite{KR}.  
It is therefore natural to make the following conjecture 
\proclaim{Conjecture 5.1}  Let $\la=M_1\La_1+\ldots+M_\ell\La_\ell$,
where $M_i\geq0$ for those $i$ such that $W_i\cong L(\La_i)$
(as a $\frak g$-module) and $M_i=0$ for the remaining $i$.  Then
$$
M_\la(q) \eql \left( \prod_{i=1}^\ell \qn{M_i}\right) \ 
h_{\wh V}(M_1,\ldots,M_\ell;q) \,.\tag{5.8}
$$
\endproclaim\medskip

Note that the condition $W_i\cong L(\La_i)$ is satisfied for all $i$
in the case of $\frak{sl}_n$, and for $i=1,n$, for $\frak{so}_{2n+1}$, i.e.,
for all the examples discussed in this paper.
We have checked Conjecture 5.1 numerically to high order for the
examples in section 4 and have a proof in special cases for
which other concise formulas for $M_\la(q)$ are known (see, in particular,
\cite{Ki1}).  Special cases of this conjecture have also appeared in
\cite{BS4}.  A general proof would clearly require a better understanding 
of the geometry of affinized flag varieties.
It is conceivable that the conjecture can be lifted to all possible
choices of $M_i$ by repeating the analysis of this paper to the 
affinized coordinate ring of a variety defined by an ideal in
$\CC[x_a^{(i)}]$ where the coordinates now correspond to a basis 
of the $U_q(\wh{\frak g})$-module $W_i$, rather than $L(\La_i)$.\pars

A different, but possibly related, relation between modified Hall-Littlewood
polynomials and the geometry of flag varieties was observed in 
\cite{HS,Sh} (see also \cite{Ki2}). \pars

The results of this paper can also be used to obtain explicit quasi-particle 
type expressions for the characters of the integrable highest weight modules
of affine Lie algebras $\whg$, in particular of $\wh{\frak{sl}_n}$ and 
$\wh{\frak{so}_5}$ in view of our results in section 4.  
These characters can be written as \cite{NY2,Ya,HKKOTY}
$$
\text{ch}_\la(q) \eql \sum_{\mu=M_1\La_1+\ldots+M_\ell\La_\ell}\ 
{1\over \prod_{i=1}^\ell \qn{M_i} } \ M_{\la\mu}^{(k)}(q) M_\mu(q) \,,
\tag{5.9}
$$
where $\la$ is the highest weight of the integrable $\whg$-module and
$M_{\la\mu}^{(k)}(q)$ a certain `level-$k$ restriction' of the polynomials
(5.2) (see \cite{BMS,DKKMM,Ki1,HKKOTY}, and references therein, for explicit 
expressions in the case of $\frak{sl}_n$, and 
\cite{BS4} for $\frak{so}_5$).
In fact, by using (5.9), the characters obtained this 
way will be of the ``Universal Chiral Partition Function'' (UCPF) form
which was recently argued to be a universal expression for the (chiral)
characters of any conformal field theory \cite{BM}.  This form reads  
$$
\text{ch}_\la(q) \eql \sum_{m_1,\ldots, m_n\geq0 \atop \text{restrictions} }
 q^{ {1\over2} \bm\cdot \bG \cdot \bm - {1\over2} 
\bA \cdot \bm}\prod_a \qbinom{ ((\id-\bG)\cdot \bm + {\bu\over2})_a}{m_a} \,,
\tag{5.10}
$$
where $\bG$ is an $n\times n$ matrix and $\bA$ and $\bu$ are certain 
$n$-vectors.  Both $\bA$ and $\bu$ as well as the restrictions on the
summations over the quasi-particle numbers $m_a$ will in general depend
on the sector $\la$, while $\bG$ will be independent of $\la$.\pars

Another interesting connection of modified Hall-Littlewood polynomials
to affine Lie algebra characters was observed (and
proved in special cases), in \cite{Ki1,NY1,KMOTU, KKN,HKKOTY,BS4} for
$\frak{sl}_n$ and in \cite{BS4} for $\frak{so}_n$.  It turns out that
these characters can often be obtained from $M_\la(q^{-1})$ in a
`large $\la$-limit', i.e., in a limit where one of the Dynkin indices
of $\la$ tends to infinity while the others are kept fixed.  This
observation is intimately related to taking the TBA limit in the
integrable spin chain underlying the definition of $M_\la(q)$
\cite{KR}. \pars

The connection of affinized projective varieties with quasi-particles
in conformal field theory, which in fact motivated this research,
arises as follows.  Quasi-particles in conformal field theory
correspond to intertwiners (`Chiral Vertex Operators' or CVOs) between
modules of the Chiral Algebra.  The degrees of freedom in these CVOs
can be separated, at least heuristically, in terms of pseudo-particles
that generate the collection of possible `fusion paths' for the CVOs
(i.e., the sequence of modules between which the CVOs intertwine) and
physical particles whose degrees of freedom can be interpreted as the
coordinate ring of an affinized projective variety; the ideal
corresponds precisely to the null-states in the physical
quasi-particle Fock space.  [In (5.10) pseudo-particles correspond 
to $u_a<\infty$ while physical particles have $u_a=\infty$.]
This paper thus gives an effective
technique to compute the contribution of the physical particles to the
conformal field theory characters.  It is hoped that similar
techniques may be applied to extract the pseudo-particle contribution.
We refer to \cite{BCR} for a more detailed exposition of this
connection, a more explicit discussion of the UCPF form of the various
affine Lie algebra characters and an application to the exclusion
statistics satisfied by these quasi-particles.  The results of
\cite{BCR} can be taken as further evidence for the correctness of
Conjecture 5.1.

% -------------------------------------------------------------------------

% -------------------------------------------------------------------------
\head Acknowledgements \endhead

P.B.\ is supported
by a \qeii\ research fellowship from the Australian Research Council.
The various Gr\"obnerbasis computations were performed with the help
of Mathematica$^{\text{TM}}$.

% -------------------------------------------------------------------------

% -------------------------------------------------------------------------
\enddocument